# WORD-BASED TEXT COMPRESSION

Jan Platoš, Jiří Dvorský
Department of Computer Science
VŠB – Technical University of Ostrava, Czech Republic
{jan.platos.fei, jiri.dvorsky}@vsb.cz

## ABSTRACT

Today there are many universal compression algorithms, but in most cases is for specific data better using specific algorithm - JPEG for images, MPEG for movies, etc. For textual documents there are special methods based on PPM algorithm or methods with non-character access, e.g. word-based compression. In the past, several papers describing variants of word-based compression using Huffman encoding or LZW method were published. The subject of this paper is the description of a word-based compression variant based on the LZ77 algorithm. The LZ77 algorithm and its modifications are described in this paper. Moreover, various ways of sliding window implementation and various possibilities of output encoding are described, as well. This paper also includes the implementation of an experimental application, testing of its efficiency and finding the best combination of all parts of the LZ77 coder. This is done to achieve the best compression ratio. In conclusion there is comparison of this implemented application with other word-based compression programs and with other commonly used compression programs.

Key Words: LZ77, word-based compression, text compression

## 1. Introduction

Data compression is used more and more in these days, because larger amount of data require to be transferred or backed-up and capacity of media or speed of network lines increase slowly.

Some data types, which still increase, are text documents like business materials, documentation, forms, contracts, emails and many others.

For compression of textual data we usually use universal compression methods based on algorithms LZ77 and LZ78. However, there are also algorithms specially developed for text like PPM or Burrows-Wheeler transformation (BWT). An interesting approach to text compression is not taking this data as sequence of characters or bytes, but as sequence of words. These words may be real words from spoken language, but also sequences of characters, which fulfill some condition, e.g. character pairs. This approach is called word-based compression.

The word-based compression is not a new algorithm, but only a revised approach to the text compression. In the past, word-based compression methods based on Huffman encoding, LZW or BWT were tested. This paper describes word-based compression methods based on the LZ77 algorithm. It is focused on different variants of algorithm itself, various implementations of the sliding window algorithm and on various possibilities of output encoding. Finally, many tests were performed to compare variants of our implementation as well as other word-based or classic compression algorithms.

## 2. Word-based compression

As mentioned above, word-based compression is not a new compression method, rather a revised approach to compressed data.

Using this approach is possible only if the structure of compressed data is known. Text files have a known structure, because they are written in some language.

Spoken language has a natural structure, which goes from separate characters through syllables and words to whole sentences.

Processing text by syllables presents one problem: we need to separate the words to syllables, preferably using grammatical structures of given language. Conversely, processing text by words is very simple because words are divided by the sequence of spaces and non-alphanumeric symbols. Word dividers are usually called non-words.

Words are represented by sequences of alphabetical characters finished by spaces or other characters. Sometimes, it is suitable to count not only words, but other sequences as well, e.g. sequences of characters and numbers starting with characters such as A1033, B5 (room numbers), or sequences containing dashes, slashes or dots like F-117, AUTOEXEC.BAT, OS/2 etc.

In the case of semi-structured text like HTML or XML documents, we do not only recognize words and non-words. Tags, which represent structural marks of given markup language, are also recognized.

A disadvantage of word-based approach in comparison with the character-based approach is impossibility to determine the size of the alphabet before compression, because its size is different for any file.

## 3. LZ77 algorithm

LZ77 is a compression algorithm which was developed in 1977 by Abraham Lempel and Jacob Ziv [1]. LZ77 belongs to a group of dictionary compression algorithms, more precisely to the subgroup of algorithms with a sliding window.

Dictionary algorithms use part of compressed text as a dictionary, which is used to compress remaining text.

LZ77 algorithm uses a part of the recently encoded text, which is stored in sliding window as a dictionary. This window is divided into two sections – the encoded section and the plain section. Compression itself consists of searching for the longest sequence in the encoded section, which is equal to the text at begin of the plain text section. Then a code triplet is sent to output. The sent code consist of the position of the found sequence (or offset from end of window), the length of this sequence and the first different character in the plain text section. The entire principle is depicted on Figure 1. During compression, the window "slides" over compressed text, hence the term "sliding".

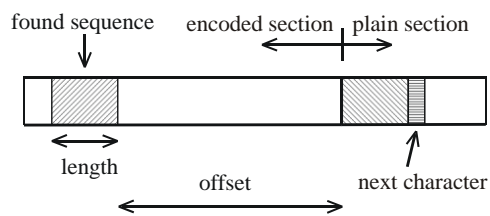

**Figure 1: LZ77 code triplet**

### 3.1 LZ77 variants

LZ77 method has many variants; most of them differ from others only in a method of storing of a code triplet. Method LZR [2] published by M. Rodeh and V. Pratt in 1981, does not constrain size of window and to store position (or offset) and length uses algorithms for large numbers encoding. Method LZSS, published by J. Storer a T. Szymanski [3] in 1982 and practically implemented by T. C. Bell [4] in 1986, stores in output only code doublet (containing offset and length of found sequence) or single characters. The decision, which variant should be used, depends on length of bit representation of code doublet and character. A signal bit, which defines, if there is stored code doublet or character, is used for correct decompression.

### 3.2 Sliding window

Implementation of sliding window is fundamental problem of any LZ77 based compression algorithm, because this implementation determine speed of whole algorithm.

Sliding window has three tasks: searching for matches, inserting new strings and removing old string. All task must be fast, but searching for matches is necessary only at the end of previous match, but inserting and removing is necessary for every symbol, hence sliding window implementation is optimized for inserting/removing and not for match searching.

Several types of sliding window, based on binary trees or hash table, were proposed in the recent years. But this proposal was intended to character based algorithm. Word-based algorithms have other requirements than character based. Therefore it was tested several variant of three base types of sliding window: binary tree, hash tables and patricia tree [5].

### 3.2.1 Binary tree
Binary tree is the oldest structure, which was used for sliding window implementation. Advantage of this structure is relatively simple implementation and very fast inserting, removing and searching. But it has one important disadvantage: though it always found longest match, thus if there are more than one occurrence of this match, then some of them is found. Other structures found always that match, which is closest to the end of window.

To improve the speed it is suitable to enlarge the number of trees. Roots of these trees are stored into array or hash table.

### 3.2.2 Hash table
Hash table is very popular structure to store dictionary and it is used in popular ZIP/GZIP program. Hash table is optimized for inserting and removing strings, not for searching and it returns always longest match closest to end of sliding window. Hash table need hashing over several characters and large table to work fast. Number of used characters determines minimal match length, which may be found. Long minimal length of match may decrease effectiveness of compression.

### 3.2.3 Patricia tree
This structure is not so frequent in sliding window implementation, but has many useful features for good effectiveness. Patricia tree keeps all advantages of digital tree but decrease memory requirements. Our proposal of implementation keeps retrieving longest match closest to the end of the window.

Enlargement of used trees is again useful for achieving of speed improvements.

### 3.2.4 Size of window, maximum length of match
Total size of sliding window and also the size of both sections differs in various implementations, e.g. program ZIP uses sliding window size 32kB and length of plain text section 256 characters. Generally true is that with larger windows can be achieved better compression.

## 3.3 Output coder
Output coder process match position, length and the first different character and it send them to the output file. Access to the match processing is based on variant of compression. In classical implementation from Lempel and Ziv match was stored in code triplet, thus all three components. In LZSS variant is stored either code doublet, contains match length and position, or only the first different symbol.

Some other improvements were developed in addition to different variants of LZ77 algorithms described above.

### 3.3.1 Lazy Evaluating
The point of lazy evaluating is simple. It consists in that, the first match found is not coded immediately, but it is searched for match on next symbol. If longer match was found than previous attempt, previous match is stored only as one character and window is moved by one symbol. If new match is shorter, then previous match ism stored as code doublet and window is moved by the length of found sequence.

### 3.3.2 Shortest path encoding
This algorithm was developed as improvement of lazy evaluating algorithm, which has some disadvantages.

If there are matches of length 4, 5, 6, 7, 8, 1 in text, then there is stored sequence of 4 characters and then match of length 8 in output, instead of much more effective match of length 4 and match of length 8.

If there exist matches of length 3, 1, 5, 1, 1, 1, 1 in text, then there is stored match of length 3 and sequence of 4 characters (or match of length 1) to the output, instead of match of length 2 and match of length 5.

The point of shortest path encoding method is again simple. Matches are searched on every position and found lengths and positions are stored in temporary file. After processing of whole file there is found that sequence, which has shortest bit representation. This algorithm worked in linear time with regard to number of edges, because found matches create oriented tree with one root and one leaf.

### 3.4 Output encoding

The last step before storing match length, position or character into file, is effective encoding of this items.

Simplest method is direct bit storage. For every item we can determine maximum of needed bits and then we can store this item so, that can by readable in decompression.

For storing match length and position is more efficient using of large number encoding method like Fibonacci codes, Elias codes or B-Block encoding (see [6]).

Usage of entropy coders like Huffman or arithmetic coder is the most effective.

## 4 Tests

Many tests were performed to compare methods' effectiveness. As testing data were used two standard compression test files: world192.txt and bible.txt from Canterbury compression corpus [7], law.txt file, which is collection of Czech law documents, access.log – log file from proxy server during two months, rfc.txt a collection of RFC documents, and latimes.txt - complete archive of articles that were published in the Los Angeles Times in 1989 and 1990 (this file has been taken from a TREC conference collection [8], [9], [10]).

Four types of output coders (LZ77, LZSS, LZSS Lazy and LZSS Short) were used for testing. LZ77 is classical coder described in [1]; LZSS is implementation of coder described in [3]. LZSS Lazy is LZSS coder improved by lazy evaluating and LZ77 Short is LZSS coder improved by shortest path encoding algorithm.

Direct bit storage, Fibonacci encoding, B-Block encoding and Huffman encoding were tested for output encoding of offset, length and characters.

### 4.1 Speed of sliding window implementation

The first test was focused on comparison between all variant of sliding window implementation.

Three variant of binary tree was chosen for testing: classical approach with one tree and extended approach with one sub tree for every input symbol and with roots stored in hash table with 256K roots and hashing over 2 symbols. As second structure was used hash table in 4 variants with hashing over one, two, three or four symbols and with size one item per symbol, 256K item, 2M items and 16M items respectively. The last used structure was patricia tree in 3 variants, same as in binary tree.

Binary tree is designed as BT, patricia tree as PT and hash table as HT. Suffix A means using of array for every item in alphabet, suffix H means using a hash table. Suffix 1, 2, 3 or 4 is number of symbols for hashing. Results are depicted in Table 1.

| File | world192.txt | | bible.txt | |
|---|---|---|---|---|
| Size | 64 kB | 1 MB | 64 kB | 1 MB |
| Win. type | time[ms] | time[ms] | time[ms] | time[ms] |
| BT | 1406 | 1797 | 2562 | 4234 |
| BTA | 1172 | 1485 | 2328 | 4032 |
| BTH | 984 | 1172 | 2047 | 3422 |
| PT | 37422 | 99047 | 25813 | 878484 |
| PTA | 9468 | 35000 | 16235 | 824250 |
| PTH | 1547 | 2750 | 5031 | 20016 |
| HT1 | 7078 | 51312 | 14250 | 240172 |
| HT2 | 1109 | 5188 | 4515 | 120281 |
| HT3 | 984 | 2828 | 3265 | 45266 |
| HT4 | 969 | 1140 | 1937 | 7375 |

**Table 1: Speed of sliding window implementation**

All measured times are only for orientation and comparison on order level.

Binary trees are the fastest implementation for all variant, but has one big disadvantage mentioned above. The second fastest structure is hash table with hashing over 4 symbols, but minimal length of match is 4 symbols. On third place there is patricia tree with hash tables. This need only 2 symbols for hashing and is much faster than hash table with 2 symbol hashing.

Patricia tree with hash table for roots was used for all following tests as sliding window.

### 4.2 Output encoding

Three tests were performed for determining best output encoding for offset, length and characters, because all items are independent.

The best storage method for offset is Huffman encoding, but it is very slow for large window. Fortunately, B-Block encoding with base 16 times smaller than windows size achieve almost equal results, hence B-Block encoding is used for offset storing.

The best storage method for character and length is Huffman encoding again.

Using of B-Block encoding for offset storage and Huffman encoding for character and length storage will be designed as best encoding.

### 4.3 Window size and maximum match

Optimal size of sliding window and maximal length of match was searched in this test. Tests with window sizes from 4 kB to 2048kB and with lengths from 4 to 64 symbols were performed.

When direct bit method was used as output encoding, then the best result was achieved with 512 kB window size and 16 symbols length of maximum match length.

When best encoding, determined in previous test, for offset, length and character was used, then better result was achieved with larger window size and the best results was achieved with window, which size was larger than size of compressed file. Length of maximum match was the most effective, when it was set on 16 or 32 symbols.

### 4.4 Output encoder comparing

Effectiveness of output encoders was compared in this test. LZ77 is classic LZ77 encoder, which stores code triplet (offset, length, character). LZSS is base LZSS encoder, which stores code doublet (offset, length) or single character. LZSS Lazy is LZSS encoder with lazy evaluating algorithm and LZSS Short is LZSS encoder with shortest path encoding algorithm.

Two variants of this test were performed. Direct bit storage for offset, length and character was used in the first variant and so-called best encoding (see section 4.2) was used in second variant.

Coder's properties were set that: window size on 1MB, maximum match length on 16.

Results of the first variant are shown in Table 2 and Table 3. Results of the second variant are shown in Table 4 and Table 5[1].

| File: | bible.txt, direct-bit storing | | |
|---|---|---|---|
| Encoder | CS [bytes] | CR [%] | time[ms] |
| LZ77 | 1133776 | 28.01 | 20328 |
| LZSS | 1052374 | 26.00 | 20093 |
| LZSS Lazy | 988821 | 24.43 | 20031 |
| LZSS Short | 958942 | 23.69 | 19593 |

Table 2: Output coder comparing, file bible.txt, direct-bit storing

| File: | law.txt, direct-bit storing | | |
|---|---|---|---|
| Encoder | CS [bytes] | CR [%] | time[ms] |
| LZ77 | 20280924 | 31.41 | 336921 |
| LZSS | 15795427 | 24.46 | 323046 |
| LZSS Lazy | 15554905 | 24.09 | 297937 |
| LZSS Short | 15241377 | 23.60 | 384422 |

Table 3: Output coder comparing, file law.txt, direct-bit storing

When direct-bit storage was used, the best results have been achieved by LZSS Short encoder.

| File: | bible.txt, best encoding | | |
|---|---|---|---|
| Encoder | CS [bytes] | CR [%] | time[ms] |
| LZ77 | 946184 | 23.38 | 23625 |
| LZSS | 876396 | 21.65 | 23500 |
| LZSS Lazy | 866416 | 21.41 | 22484 |
| LZSS Short | 864310 | 21.35 | 22531 |

Table 4: Output coder comparing, file bible.txt, best encoding

| File: | law.txt, best encoding | | |
|---|---|---|---|
| Encoder | CS [bytes] | CR [%] | time[ms] |
| LZ77 | 15990252 | 24.76 | 537906 |
| LZSS | 13231130 | 20.49 | 490203 |
| LZSS Lazy | 13036056 | 20.19 | 471203 |
| LZSS Short | 13187228 | 20.42 | 550875 |

Table 5: Output coder comparing, file law.txt, best encoding

---

[1] CR mean compression ratio and CS mean size after compression

The best results in second variant of test have been achieved by LZSS Lazy encoder, because LZSS Short encoder was not proposed for updating model of entropic encoder during calculation of "shortest path" in graph.

Through above mentioned disability has been as the best encoder chosen LZSS Lazy encoder.

method and second-time with this method. Results are in Table 7.

Compression methods specially designed for text are marked with "-t". Text compression method in 7ZIP is based on PPM algorithm and in RAR probably too, but RAR is commercial software and detailed information about algorithms was not published.

|  | Huffword | | WLZW | | WLZ77 | |
| --- | --- | --- | --- | --- | --- | --- |
| File: | CS [bytes] | CR [%] | CS [bytes] | CR [%] | CS [bytes] | CR [%] |
| world192.txt | 686395 | 27.75 | 525205 | 21.23 | 443001 | 17.91 |
| bible.txt | 1150404 | 28.42 | 972972 | 24.04 | 866416 | 21.41 |
| law.txt | 20210941 | 31.30 | 17389900 | 26.93 | 13036056 | 20.19 |
| access.log | 44252003 | 26.07 | 19601388 | 11.55 | 16788755 | 9.89 |
| rfc.txt | 45862495 | 25.52 | 39619488 | 22.05 | 30991468 | 17.25 |
| latimes.txt | 152619478 | 30.62 | 158923990 | 31.89 | 121275213 | 24.33 |

Table 6: Comparing with other word-compression based method

|  | bible.txt | | law.txt | | latimes.txt | |
| --- | --- | --- | --- | --- | --- | --- |
| Program: | CS [bytes] | CR [%] | CS [bytes] | CR [%] | CS [bytes] | CR [%] |
| 7ZIP-t | 713741 | 17.63 | 11214471 | 17.38 | 102229081 | 20.51 |
| RAR-t | 726723 | 17.96 | 11685797 | 18.10 | 109748277 | 22.02 |
| **WLZ77** | **866416** | **21.41** | **13036056** | **20.19** | **121275213** | **24.33** |
| BZIP2 | 845623 | 20.89 | 14206832 | 22.00 | 137371338 | 27.56 |
| 7ZIP | 885104 | 21.87 | 12224454 | 18.93 | 125573610 | 25.19 |
| RAR | 954326 | 23.59 | 13382460 | 20.72 | 133896909 | 26.88 |
| GZIP | 1176645 | 29.07 | 18270683 | 28.29 | 175864812 | 35.29 |

Table 7: Comparing with standard compression programs

### 4.5 Word-based compression method comparing

Main point of this work was effectiveness of LZ77 algorithm in word-based compression. Hence LZ77 compressor (designed as WLZ77) was compared in this test with compressors based on Huffman encoding (designed as HuffWord) and LZW algorithm (designed as WLZW [11]). Results are shown in Table 6.

This test shown, that LZ77 algorithm is much better than other two algorithms. HuffWord was worse sometime till 10 percent of compression ratio.

### 4.6 Comparing with other programs

This section describes comparison between WLZ77 algorithm and compression programs, which is usually used in practice.

Compression level of all used programs was set on maximum, and if would be possible, the size of dictionary was set on 4 MB. If programs had special compression method for text files, then the test was performed twice, first-time without this

Programs based on LZ77 algorithm (GZIP, 7ZIP and RAR) are mostly worse than WLZ77 method. BZIP2, which is based on Burrows-Wheeler transformation, is some time better sometime worst. Specially designed methods based on PPM are always better, but PPM algorithm is practically symmetric. That's mean; the time of decompression is generally equal to the time of compression and then slows. LZ77 algorithm is asymmetric and it has much faster decompression than compression.

## 5 Conclusion

LZ77 algorithm is very effective in word-based compression, but its design and implementation is very difficult. Sliding window spend most of compression time for inserting, searching and removing strings. In comparison with other word-based method it is the best. In comparison with standard compression programs (i.e. mostly character based) it is still behind method based on PPM algorithm.

In future work it is possible to improve sliding window implementation, output encoders and encoding and, maybe, transform algorithm to (at least) parallel for faster work on multi core/cpu computers.

**Acknowledgement**
This work is supported by Grant of Grant Agency of Czech Republic No. 201/05/P145.